\documentclass[conference,final,twocolumn]{IEEEtran}
\usepackage{amssymb}
\usepackage{amsmath}
\usepackage{epsfig}
\usepackage{algorithm}
\usepackage{graphicx}
\usepackage{subfigure}
\usepackage{bm}
\usepackage{amsthm}
\usepackage{multirow}
\usepackage{tikz}
\usetikzlibrary{arrows.meta}
\usetikzlibrary{bending}
\usetikzlibrary{topaths}

\newtheorem{myProp}{Proposition}
\newtheorem{myProf}{Proof}

\allowdisplaybreaks

\begin{document}

\title{Low PAPR Pilot for Delay-Doppler Domain Modulation}
\author{\IEEEauthorblockN{Pu~Yuan}
\IEEEauthorblockA{vivo Communication Research Institute, Beijing, China 100015\\} 
Email: pyuan2@e.ntu.edu.sg}
\maketitle

\begin{abstract}
This paper studies the low peak-to-average power ratio (PAPR) pilot design in delay-Doppler domain modulation. We adopt a sequence based pilot design to mitigate the high PAPR brought by the conventional pulse pilot. We develop a two-stages channel estimation algorithm composed of path identification and channel coefficient estimation. We provide the quantitative analysis on the channel estimation error model, and shows the principle in designing sequence pilot in delay-Doppler domain. Experiment results show that the proposed scheme maintains a relatively low PAPR in time domain samples, while the channel estimation performance approaches the ideal channel estimation in limited-Doppler-Shift channel model. 
\end{abstract}


\section{Introduction}
\subsection{Motivation}
Delay-Doppler domain modulation (DDM) is an emerging technique which is potentially to be a candidate feature of the next generation wireless communication systems. Different from the CP-OFDM or DFT-s-OFDM in the contemporary system, the DDM modulates the QAM symbols in the delay-Doppler domain instead of the time-frequency domain. The DDM based system is enabled to perform the signal analysis and processing in the delay-Doppler domain, and benefits from the time-invariant property of the delay-Doppler domain channel. One well known variant of the DDM is the orthogonal time frequency space (OTFS) modulation \cite{Hadani2017otfs}, which employes an ISFFT before the OFDM modulator in order to implement the resource mapping in the delay-Doppler domain.

\begin{figure}
\centering
        \includegraphics[width=0.45\textwidth]{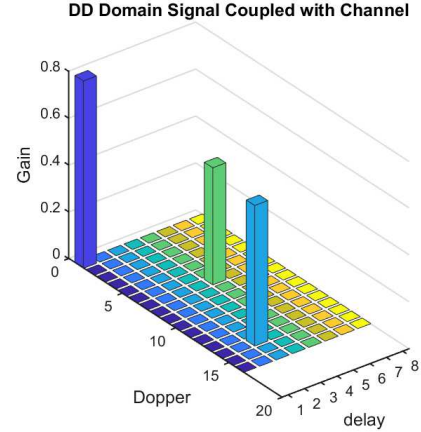}
        \caption{Illustration of delay-Doppler domain channel response. The three paths here each with the channel responses $<0.8, 1, 1>$, $<0.6, 3, 14>$ and $<0.5, 5, 7>$.}
    \label{fig:fig0}
\end{figure}

A typical delay-Doppler channel response is illustrated in Fig. \ref{fig:fig0}. The channel response couples with the transmitted signal in a doubly-convolution manner, results in offsets in both dimensions of the delay-Doppler plane. Therefore, the received signal in delay-Doppler domain is the circulant shifted versions of the transmitted one, and multiplied by  complex valued channel coefficients. If we denote $h_i$ as the channel coefficient, $\tau_i$ as the delay, $\nu_i$ as the Doppler, then a propagation path $p_i$ can be uniquely defined by a set $<h_i,\tau_i,\nu_i>$. 

The physical meaning of these parameters are the follows. The channel coefficient is a mixture of the propagation loss, including the path loss, absorption loss, shadowing, etc., and the phase changed by the reflectors. The delay is solely contributed by the distance traveled by the radio. The Doppler is also a mixture effort caused by the moving state, mainly the direction and velocity, of the communicating nodes and the position of the reflectors. Mathematically, the delay-Doppler domain channel can be modeled as,

\begin{align}
h(\tau, \nu) = \sum^{P}_{p = 1}{h_p \delta(\tau - \tau_p) \delta(\nu - \nu_p)}.
\label{eq:dDCh}
\end{align}

\begin{figure}
\centering
        \includegraphics[width=0.45\textwidth]{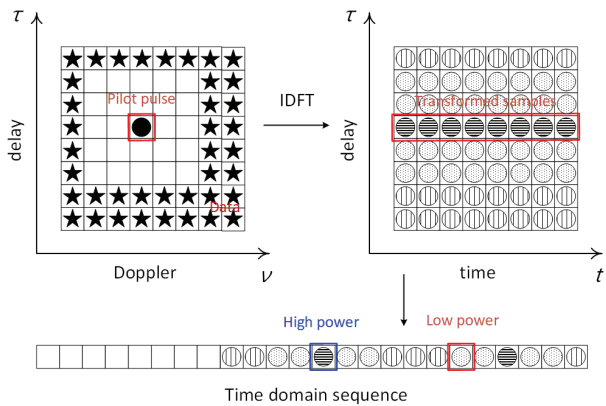}
        \caption{Conventional pulse pilot design suffers the high PAPR of the time domain samples.}
    \label{fig:fig1}
\end{figure}

Inspired by (\ref{eq:dDCh}), a straightforward way to estimate the channel response in delay-Doppler domain is invented. By transmitting a high power pulse, while reserves enough guard space surrounding the pilot, the potential mutual interferences between the pilot and data after channel coupling can be successfully avoided \cite{Raviteja2019embedded}. After delay-Doppler resource mapping, the pilot and data are converted to delay-time domain together and then vectorized to time domain samples to be sent. This approach is illustrated in Fig. \ref{fig:fig1}. At the receiver side, the received time domain samples are reverted to the delay-Doppler domain. The pulse pilot after channel coupling experiences delay and Doppler offsets within the guard area, i.e., several shifted copies with degraded power can be detected, thus the channel estimation can be done by simple power detector.

However, as illustrated in Fig. \ref{fig:fig1}, the high power pulse pilot causes imbalanced power distribution in each delay tap. Suppose the pulse pilot locates in delay tap $l$, then after IDFT, the average power of the delay-time domain samples in delay tap $l$ will be significantly larger than those in other delay taps. Consequently, periodical peaks are  observed in the time domain samples. This characteristic is unfriendly for hardware implementation. 

\subsection{Related Works}
Most of the research works in DDM focus on the OTFS \cite{Hadani2018orthogonal} waveform. To fully explore the diversity gain brought by the cross symbol spreading of the OTFS, non-linear iterative message passing algorithm (MPA) detector has been proposed in \cite{Raviteja2018interference}. Benefited from the sparsity of delay-Doppler domain channels, the performance and complexity are well trade-off. To further reduce the symbol detection complexity, the method in \cite{Raviteja2018lowcomplexity} sacrifices the spectrum efficiency by inserting Zero-padding (ZP) in the delay Doppler domain symbols to mitigate the inter Doppler interference, then low complexity MRC detector can be utilized for symbol detection. Further improvement based on the ZP-OTFS framework is presented in \cite{Qi2022iterative}, where iterative MMSE equalizer is shown to outperform the RAKE detector. 

Equipped by multiple antennas, the performance of DDM can be further improved thanks to the extra degree of freedom brought by MIMO. In \cite{wang2021transmit}, the transmit diversity is achieved by applying space time code to OTFS in multiple antennas system. In \cite{Liu2020uplink}, aided by the up-link reference signal, multiple transmit antennas perform the transmit beam-forming to further decompose the delay-Doppler domain channel into overlapped angular-sliced-delay-Doppler channels, which with more sparsity and significantly reduce the channel estimation and symbol detection complexity. Similarly, in \cite{Shan2021lowcomplexity} the receive beam-forming is utilized to increase the sparsity of angular-sliced channels, which enables to use a low-complexity detection scheme with smaller pilot overhead. In these works, an extra the channel estimation gain is collected from multiple antennas. 
 
The technique focus of this article, i.e., the PAPR issue in DDM also receives a few contributions. In general, the PAPR issue can be categorized into two types. The first type is the PAPR of the original DDM waveform. The PAPR analysis of OTFS under different pulse shaping scheme has been discussed in \cite{Surabhi2019peak}, the author shows that in general the PAPR of OTFS is lower than the OFDM and GFDM. The PAPR reduction technique has been discussed in \cite{Naveen2020peak}, where the author employs a MuLaw companding technique to refine the waveform while slightly sacrificing the symbol detection performance. While these contributions mainly discuss the properties and refinement of the original OTFS waveform with data symbols only, it lacks discussion on the PAPR issue caused by the high power pulse pilot, which we defined as the second type problem. 

The main idea of this work is to employ a sequence pilot instead of pulse pilot to balance the power distribution in each delay tap of the delay-Doppler resource grid. The sequence pilot spread along the delay dimension and leaves guard space in Doppler dimension only. At the receiver side, the pilot and its doubly-shifted versions are not guaranteed to be orthogonal due to lack of delay dimension guard space. Hence interference-resilience PN sequence should be adopted for path identification and channel gain estimation. Note that there are also a few contributions mentioned the PN sequences based pilots in the delay-Doppler domain \cite{Liu2020uplink}\cite{Shen2019channel}, however their purposes are to reduce the pilot overhead rather than mitigate the PAPR, which leads to a block-wise mapping pattern with guard space in both dimension. The PAPR problem are still there, i.e., it is mildly reduced but not eliminated. To the best of the author's knowledge, this is the first contribution try to reduce the PAPR of DDM from the pilot design's perspective.

The remaining of the paper is organized as follows. Section II introduces the proposed pilot design and the detection algorithm. Section III presents the performance analysis. Section IV shows the numerical results. Section V concludes the paper.

\section{Low PAPR Pilot Design}
In this section, we first brief how the channel couple with the sequence pilot, then give in details the pilot pattern and sequence design. We use bold capital letters for matrices, bold lowercase letters for vectors and italic lowercase letters for scalars.
\subsection{Channel Coupling on Sequence Pilot}
\begin{figure}
\centering
\subfigure[The sequence before transmit.]{\includegraphics[width=0.45\textwidth]{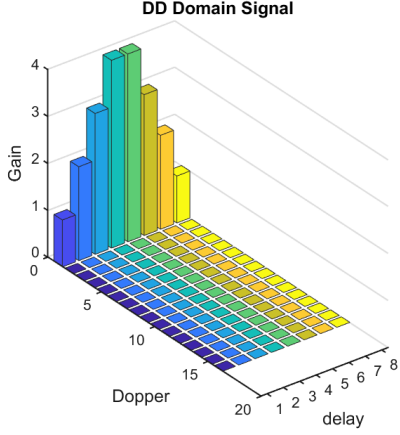}}\\
\subfigure[The sequence after channel coupling.]{\includegraphics[width=0.45\textwidth]{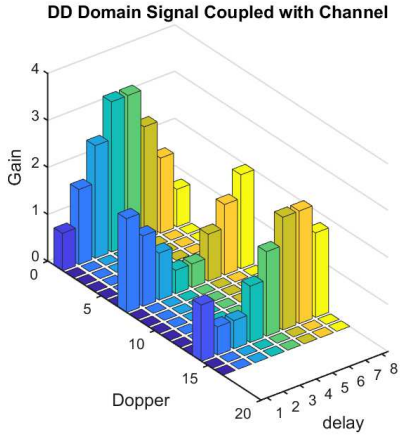}}\\
  \caption[]{Illustration of channel coupling on sequence pilot in delay-Doppler domain. We use different magnitude to differentiate different entries of the sequence.}
	{\label{fig:fig3}} 
\end{figure}

Suppose we have a sequence pilot located at Doppler tap $1$ and occupied all delay taps. The channel coupling effect on transmitted sequence in the delay-Doppler domain is illustrated in Fig. \ref{fig:fig3}. Fig. \ref{fig:fig3}(a) is the original sequence locates in Doppler tap $1$. After coupled by the channel in Fig. \ref{fig:fig0}, the received signals with cyclic-shifts in delay and offsets in Doppler appear in the delay-Doppler plane, as illustrated in Fig. \ref{fig:fig3}(b). The signal experiences no delay or Doppler offset through the first path. The signal experience $3$-cyclic-shift in delay and $14$-offset in Doppler through the second path. The signal experience $5$-cyclic-shift in delay and $7$-offset in Doppler through the third path. 
%

\begin{figure}
\centering
        \includegraphics[width=0.45\textwidth]{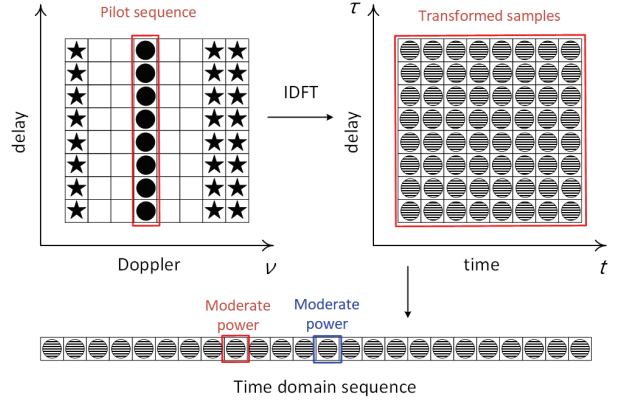}
        \caption{Sequence based pilot balances the power allocation among delay taps, therefore mitigates the PAPR of time domain samples.}
    \label{fig:fig2}
\end{figure}

The design philosophy of the sequence pilot follows the first principle. Since the high PAPR  is caused by imbalanced power distribution among delay taps by using pulse pilot, we instead use a sequence pilot based design without violating the balanced inter-delay power allocations. The pilot spreads along all the delay taps with equally powered elements. Consequently, the magnitude variation of the time domain samples are small. The corresponding mapping is illustrated in Fig. \ref{fig:fig2}. 

The following work is the choice of the sequence pilot. As there is no guard space in delay taps, the received signal may suffered from inter-delay interference (IDI) , hence finding a proper sequence resilience to the IDI is important. Usually we tend to choose the sequences with the property that has high auto correlation but low cross correlation when applying correlation based signal detection. Here we adopt the M-sequence as the sequence pilot \cite{Golomb1967shift}. It has a nice property that, for any length-$L$ M-sequence, the auto correlation equals to $L$ and the cross correlation with its cyclic-shifted version is $-1$. 

\subsection{Channel Estimation}

Here algorithm $1$ is provided for channel estimation based on proposed scheme. Generally speaking, the channel estimation composes of two stages. The first stage is the path identification achieved by correlating the known sequence with the received signal, the output is a set of resolvable paths uniquely correspond to a set of real valued delay-Doppler pairs. The second stage is the channel coefficient estimation based on the output of stage-$1$. 
\begin{algorithm}
Step - 1) \textit{Path Identification:} 
\begin{itemize}
\item[a)] A detection matrix $\mathbf{P_{d}}$ is constructed based on sequence pilot $\mathbf{s}$, where,
\begin{equation}
\mathbf{P_{d}} = [\mathbf{s}^T_{\mathcal{C}({0})},\mathbf{s}^T_{\mathcal{C}({1})},...,\mathbf{s}^T_{\mathcal{C}({N-1})}]^T, \nonumber 
\label{eq:pd0}
\end{equation}
and $\mathbf{s}^T_{\mathcal{C}({i})}$ is the $i$-cyclic-shift of sequence $\mathbf{s}$. 

\item[b)] $\forall k$ within the detection area, denote the samples from $k_{th}$ Doppler tap as $\mathbf{r}_{kM}$, calculate the correlation vector, 
\begin{equation}
\mathbf{v}_{kM} = \mathbf{P_{d}}\mathbf{r}_{kM} = [v_1, v_2, ...,v_{N-1}], \nonumber
\label{eq:pd1}
\end{equation}

\item[c)] 
If a sub-set of elements in $\mathbf{v}_{kM}$, say $\mathbf{v}_{d}$, satisfies,
\begin{equation}
\mathbf{v}_{d} = \forall v_l \in \mathbf{v}_{kM}, v_l>\beta, \nonumber
\label{eq:pd2}
\end{equation}
where $\beta$ is a pre-defined threshold. Then all delay paths with Doppler $k$ is identified, i.e., a set of delay-Doppler pair $<l_p,k_p>$ is obtained.

\end{itemize}

Step - 2) \textit{Channel Coefficient Estimation:}
\begin{itemize}
\item[a)]  If there is only one delay tap $l_p$ corresponds to one Doppler tap $k_p$, then a simple point-wise division will solve the path gain vector,
\begin{equation}
\mathbf{h}_{p} = \left[\frac{\mathbf{r}_{kM}(1)}{\mathbf{s}_{p}(1)},\frac{\mathbf{r}_{kM}(2)}{\mathbf{s}_{p}(2)},... ,\frac{\mathbf{r}_{kM}(M-1)}{\mathbf{s}_{p}(M-1)}\right], \nonumber
\label{eq:pd3}
\end{equation}
averaging $\mathbf{h}_{p}$ to get the estimated channel coefficient of path $p$ as $\bar{h}_{p}$.

\item[b)] If there are multiple delay taps correspond to one Doppler tap $k_p$, then we can solve a set of channel coefficients of all paths by constructing a system of linear equations, which is discussed in equation (\ref{eq:ag1})  to (\ref{eq:ag6}) in Section III.
\end{itemize}
\label{alg:algc}
\caption{Example of Sequence Pilot Based Channel Estimation }
\end{algorithm}

\section{Performance Analysis}
In this section we provide the performance analysis on our proposed channel estimation scheme. We denote a $N\times M$ matrix $\mathbf{A}^{+}$ as the \textit{pseudo inverse} of a $M\times N$ matrix $\mathbf{A}$, where,
\begin{equation}
\mathbf{A}^{+}\mathbf{A}=\mathbf{I}_N.
\label{eq:ag0}
\end{equation}

The sequence pilot will experience offsets in the Doppler dimension, and the cyclic shifts in the delay dimension. If there are multiple delays correspond to one Doppler, then different cyclic shifted versions of the transmitted signal will overlap at one Doppler tap, i.e., they may mutually interfere. In this section we will discuss if this interference will negatively impact the channel coefficient estimation.

The channel coefficient estimation in stage-$2b)$ can be equivalent to solving a system of linear equations. Note that only approximately solutions can be obtained in presence of noise. Suppose there are in total $P$ delay paths in Doppler tap $k$, $\forall p\in[1,P]$, there is a delay value $l_p$ and a channel coefficient $h_p$. The received pilot signal in Doppler tap $k$ is given by,
\begin{equation}
\mathbf{y} = \sum^P_{p=1}{h_p\mathbf{x}_{\mathcal{C}({l_p})}+\mathbf{w}}.
\label{eq:ag1}
\end{equation}

In (\ref{eq:ag1}), $\mathbf{x}$ is the sequence pilot we transmit, $\mathbf{x}_{\mathcal{C}({l_p})}$ corresponds to the $l_p$-cyclic-shift of $\mathbf{x}$ which is shown in (\ref{eq:ag2}) and the noise term is a complex Gaussian where $\mathbf{w}\sim \mathcal{CN}(0,\sigma^2)$.
\begin{equation}
\mathbf{x}_{\mathcal{C}({l_p})} = 
  \begin{tikzpicture}[
    baseline=(a0.base),
    inner sep=0pt,
    line cap=round,
    >={Computer Modern Rightarrow[bend]},
  ]
    \node (a0) {$[x_{M-l_p}$};
    \node[anchor=base west] (a1) at (a0.base east)
      {${}\to \dots \to x_{M-1}\to x_0 \to \dots \to{}$};
    \node[anchor=base west] (ak) at (a1.base east)
      {$x_{M-1-l_p}]$};
    \draw[
      ->, 
      transform canvas={yshift=.3em},
    ] (ak) to[bend right, looseness=.6] (a0);
  \end{tikzpicture}
	\label{eq:ag2}
\end{equation}

Note that in the stage-$2$ of channel estimation, (\ref{eq:ag1}) is solvable since $\mathbf{y}$ is the received signal in particular Doppler tap and all $\mathbf{x}_{\mathcal{C}({l_p})}$ are found in stage-$1$. Equation (\ref{eq:ag1}) can be rewritten as,
\begin{equation}
\mathbf{y} = \mathbf{X}\mathbf{h}+\mathbf{\tilde{w}},
\label{eq:ag3}
\end{equation}
where $\mathbf{\tilde{w}}\sim \mathcal{CN}(0,\sigma^2)$ as it is linear combinations of $\mathbf{w}$. 

The matrix $\mathbf{X}$ is a $M\times P$ matrix given by,
\begin{equation}
\mathbf{X} = [\mathbf{x}^T_{\mathcal{C}({l_1})},\mathbf{x}^T_{\mathcal{C}({l_2})},...,\mathbf{x}^T_{\mathcal{C}({l_P})}].
\label{eq:ag4}
\end{equation}

Since the columns of $\mathbf{X}$ are the cyclic shifted versions of a PN sequence, then $\mathbf{X}$ is naturally full rank, i.e., it is pseudo invertible. Therefore we can approximately solve $\mathbf{h}$ according to (\ref{eq:ag2}),
\begin{equation}
\mathbf{\tilde{h}} = \mathbf{X}^{+}\mathbf{y}.
\label{eq:ag5}
\end{equation}

The estimation error of channel coefficients is thereby given by,
\begin{equation}
\mathbf{e}=\mathbf{h}-\mathbf{\tilde{h}} \\
= \mathbf{X}^{+}\mathbf{\tilde{w}}.
\label{eq:ag6}
\end{equation}

Based on above relationship, we have the following proposition.
\begin{myProp}
The average channel estimation error $\mathbf{e}$ is a random variable satisfying $\mathbf{e} \sim \mathcal{CN}(0,\epsilon^2)$, where $\epsilon^2 = \frac{MP-(P-1)^2}{MP(M-P+1)}\sigma^2$, if M-sequence based pilot is used.
\label{prop:OCFG1}
\end{myProp}

\begin{myProf}:
Write (\ref{eq:ag6}) in vector form, we obtain,
\begin{align}
[\mathbf{e}_1,\mathbf{e}_2,...\mathbf{e}_{M-1}]^T = [\mathbf{\hat{x}}_1^T,\mathbf{\hat{x}}_2^T,...\mathbf{\hat{x}}_{M-1}^T]^T\mathbf{\tilde{w}}.
\label{eq:ag7}
\end{align}

It is noted that for each $i$ we have,
\begin{align}
\mathbf{\mathbf{e}}_i = \mathbf{\hat{x}}_i\mathbf{\tilde{w}},
\label{eq:ag8}
\end{align}
since $\forall i, \mathbf{e}_i$ is the linear combination of random variables $\mathbf{\tilde{w}}$, thus the expectation of $\mathbf{e}_i$ is also zero.

Consequently, the self-covariance matrix of $\mathbf{e}$ can be calculated as,
\begin{align}
E[\mathbf{e}\mathbf{e}^T] & = E[\mathbf{X}^{+}\mathbf{\tilde{w}}\mathbf{\tilde{w}}^T(\mathbf{X}^{+})^T] \nonumber  \\
& = \mathbf{X}^{+}E[\mathbf{\tilde{w}}\mathbf{\tilde{w}}^T](\mathbf{X}^{+})^T
 = \mathbf{X}^{+}(\mathbf{X}^{+})^T\mathbf{I}\sigma^2 \nonumber  \\
& = (\mathbf{X}^{T}\mathbf{X})^{-1}\mathbf{I}\sigma^2.
\label{eq:ag9}
\end{align}

Recall that the column vectors of $\mathbf{X}$ are $\mathbf{x}_{\mathcal{C}({l_p})}$, which are M-sequences, then it is easy to verify that $\mathbf{X}^T\mathbf{X}$ is a $P\times P$ symmetric matrix of the following form,
\begin{equation}
\mathbf{X}^T\mathbf{X} = 
\left[
\begin{array}{cccc}
M & -1 & ... & -1 \\
-1 & M & ... & -1 \\
      .         &       .        & .\ \    &       .          \\
      .         &       .        & \  . \  &       .          \\
-1 & -1 & ... & M \\
\end{array}
\right].
\label{eq:ag10}
\end{equation}

Hence a spectral decomposition \cite{Halmos1958finite} can be applied to $\mathbf{X}^T\mathbf{X}$, where,
\begin{align}
\mathbf{X}^T\mathbf{X} & = \mathbf{Q}^{-1}\bm{\Lambda}\mathbf{Q} \nonumber \\
& = \mathbf{Q}^{-1}
\left[
\begin{array}{cccc}
M-(P-1) & 0 & ... & 0 \\
0 & M & ... & 0 \\
      .         &       .        & .\ \    &       .          \\
      .         &       .        & \  . \  &       .          \\
0 & 0 & ... & M \\
\end{array}
\right] 
\mathbf{Q}.
\label{eq:ag11}
\end{align}

Substitute (\ref{eq:ag11}) into (\ref{eq:ag9}), we have,
\begin{align}
E[\mathbf{e}\mathbf{e}^T]
& = \mathbf{Q}^{-1}\bm{\Lambda}^{-1}\sigma^2\mathbf{Q} \nonumber \\
& = \mathbf{Q}^{-1}
\left[
\begin{array}{cccc}
\frac{\sigma^2}{M-P+1} & 0 & ... & 0 \\
0 & \frac{\sigma^2}{M} & ... & 0 \\
      .         &       .        & .\ \    &       .          \\
      .         &       .        & \  . \  &       .          \\
0 & 0 & ... & \frac{\sigma^2}{M} \\
\end{array}
\right] 
\mathbf{Q}.
\label{eq:ag12}
\end{align}

Then the average power of all random variables in $\mathbf{e}$ is obtained by averaging the diagonal elements in $\bm{\Lambda}^{-1}\sigma^2$,
\begin{equation}
\epsilon^2 = \frac{MP-(P-1)^2}{MP(M-P+1)}\sigma^2.
\label{eq:ag13}
\end{equation}
\qed
\end{myProf}

Proposition \ref{prop:OCFG1} gives us an insight on how the sequence pilot design can quantitatively impact the channel estimation performance. First of all, different types of sequences will lead to different channel estimation errors. In this work, since the M-sequence is adopted, then the non-diagonal entries of $\mathbf{X}^T\mathbf{X}$ are all minus one, which impacts the eigenvalues in the manner of (\ref{eq:ag11}). If other types of sequence are used, it would in other forms. It is observed that the distribution of channel estimation error will approach $\mathcal{CN}(0,\sigma^2)$ as the absolute value of the non-diagonal entries of $\mathbf{X}^T\mathbf{X}$ approaches zero.

Secondly, the observation on how the two variables $M$ and $P$ impact the equivalent SNR in channel estimation is also useful. Suppose now we are doing a comparison of proposed pilot with the conventional pulse one. We should apply the same total power constraint $\varpi$ for fairness. Hence for the pulse pilot, the noise power is $\sigma^2$, then the SNR of the pulse pilot is given by,
\begin{equation}
SNR_{P} = \frac{\varpi}{\sigma^2}, 
\label{eq:ag14}
\end{equation} 
which is considered as a benchmark. With the results in Proposition \ref{prop:OCFG1}, the equivalent SNR of proposed channel estimation scheme can be written as,
\begin{equation}
SNR_{S} = \frac{\varpi/M}{\epsilon^2} = \frac{SNR_{P}}{1+\frac{P-1}{P(M-P+1)}}, 
\label{eq:ag15}
\end{equation} 
where $\frac{P-1}{P(M-P+1)}$ is the deviation of $SNR_{S}$ from $SNR_{P}$. It is positive, decreases with $P$ and increases with $M$. This quantitative result confirms our intuitive thoughts. First, the longer the sequence, the less channel estimation error. Second, the less mutually interfering pilots in the same Doppler tap, the less the channel estimation error. When $P=1$, i.e., there is no mutually interfering signal, then the channel estimation error only depends on noise power.

\section{Numerical Results}
In this section we provide the numerical results of the performance of the proposed sequence pilot based scheme with the conventional pulse pilot. The channel model we adopted is the limited Doppler-shift channel (LDSC) model, which is commonly adopted in the ISAC system research. Consider a delay-Doppler domain frame of size $N\times M$, where $N=32$ and $M=64$ are the Doppler and delay, respectively. The delay profile of the channel is $[0, 1, 2, 3, 4, 5]$. The Doppler profile is $[0, 1, 2, 3, 4, 5]$, which contains rather high Doppler. The channel coefficient is calculated based on the aforementioned delay profile with a randomly draw power profile generated in each frame. The modulation and coding scheme is QPSK and LDPC, $r=0.5$. 

For the sequence pilot, we use a M-sequence generated by the primitive polynomial $[1\ 0\ 0\ 0\ 0\ 1]$. The sequence is of length $63$ and a dummy symbol is attached to fulfill the $64$ delay taps. The guard band in Doppler occupies $10$ symbols on each side. The sequence pilot is power boosted by borrowing the unused power from the guard symbols, therefore the total power of every entries of the sequence pilot is about $31dB$ over one data symbols. We adopt conventional pulse pilot with the same resource overhead and power to do a fair comparison. The pulse pilot adopts a power detector follow the same line as which in \cite{Raviteja2019embedded}.

\begin{figure}
    \centering
        \includegraphics[width=0.45\textwidth]{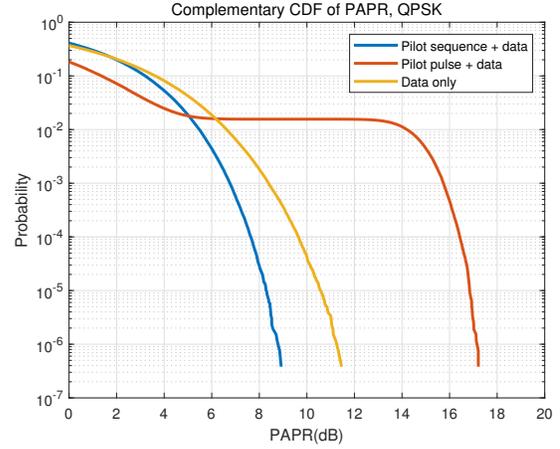}
        \caption{The PAPR comparison based on CCDF. The proposed scheme outperforms the pulse pilot while slightly better than data only, since the guard space in Doppler smooth the power distribution.} 
    \label{fig:Num1}
\end{figure}

Fig. \ref{fig:Num1} shows the PAPR performance of the proposed scheme compared with the conventional ones. We observe that the PAPR of the time domain samples in our design are bellow $9dB$, there are about $2\%$ of the simulated PAPR points are between $5dB$ and $9dB$. The result is even slightly lower than the benchmark with data only, which contains about $2\%$ of the simulated PAPR points between $6dB$ and $12dB$. In contrast, more than $2\%$ of the simulated PAPR points locate between $5dB$ and $17dB$. Such high PAPR will cause great trouble in hardware implementation.

\begin{figure}
\centering
\includegraphics[width=0.45\textwidth]{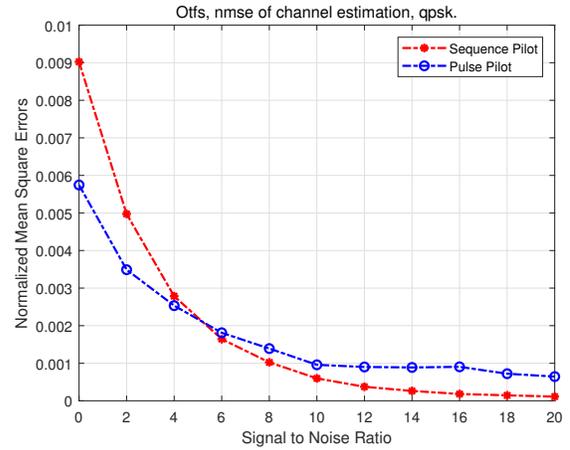}
  \caption[]{\label{fig:Num2}{} The channel estimation performance of proposed scheme.}
\end{figure}

The channel estimation accuracy is shown in Fig. \ref{fig:Num2} in terms of NMSE. It is observed that the channel estimation error of proposed scheme approaches zero when SNR is greater than $15dB$. The performance of the sequence pilot is worse than the pulse pilot when SNR is less than $5dB$, and better in other SNR range. It is observed that the main contribution to the channel estimation is the false detection of the path. The propose sequence pilot obtains extra diversity gain by delay dimension spreading, therefore more robust in path identification.

\begin{figure}
\centering
\includegraphics[width=0.45\textwidth]{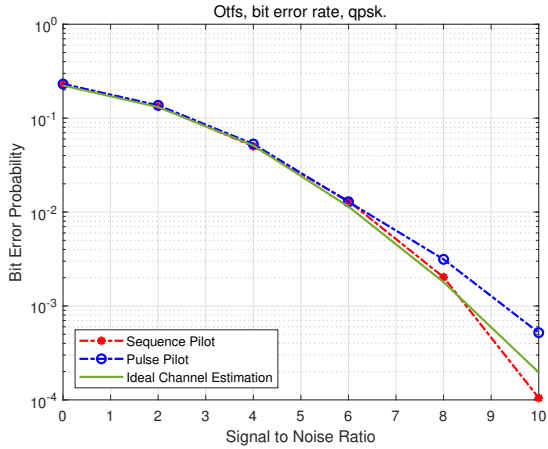}
  \caption[]{\label{fig:Num3}{} The symbol detection performance comparison based on BER. The ideal channel estimation for MPA seems not that perfect since deep faded paths contribute negative effect on the symbol detection.}
\end{figure}

Fig.\ref{fig:Num3} shows the symbol detection performance based on the channel estimation of proposed scheme and conventional pulse pilot. The MPA proposed in \cite{Raviteja2018interference} is adopted as the symbol detection algorithm. The benchmark is the ideal channel estimation where the receiver directly utilizes the channel matrix generated in simulation. It is observed that the proposed scheme out performs the conventional one start from SNR equals to $5dB$, and about $1.5dB$ better when bit error rate (BER) is $0.1\%$. This gain comes from the extra diversity collected by the sequence pilot spreading all over the delay dimension.

An interesting observation is that although the curves locates closely throughout the SNR range, at some point the proposed scheme perform slightly better than the benchmark. This phenomenon can be explained as follows. The adopted channel model will randomly generate some paths in ultra deep fading, results in the power of the signal transmitted in these paths flooded below the noise level. While using the ideal channel estimation, these deep faded paths are also counted in the symbol detection, which makes a negative contribution as the input to the detector is almost noise. In contrast, when adopting the realistic channel estimation, the deep faded paths can not be even identified, then only the received signal with good quality are considered. This result may inspire us to improvement the work mechanism of the MPA by mitigating the negative impact of the 'bad' path. 

The consideration of the proposed scheme in continuous Doppler-shift channel (CDSC) will be studied in the follow up work. The performance is expected to be degraded by the quantization error of the fractional Doppler, which is hardly captured using current channel estimation algorithm.

\section{Conclusion}
The low PAPR pilot design for the DDM waveform and the corresponding detection scheme are studied in this work. We have proposed a sequence pilot spreading along the delay dimension to balance the power distribution among delay taps. The error model of the channel estimation has been analyzed, based on which a low complexity channel estimation algorithm is provided. The PAPR of the proposed pilot design is significantly smaller than the conventional pulse pilot, which is comparable to the signal contains data only. We provide a methodology to quantitatively analyze the relationship between the channel estimation error and the pilot length and channel matrix. Benefited from the spreading gain in the delay dimension, the channel estimation performance outperforms the pulse pilot design under the LDSC.  
Future work includes the expansion of the proposed design to the CDSC which is more realistic in communication systems.

\section*{Acknowledgement}
Thanks to my colleague Dr. Junjie Tan for the helpful comments to the manuscript.

\bibliographystyle{IEEEtran}
\bibliography{fmtc_TWC}
\end{document}